\documentclass
[twocolumn,showpacs,preprintnumbers,amsmath,amssymb]
{revtex4}
\usepackage{graphicx}
\usepackage{dcolumn}
\usepackage{bm}
\usepackage{here}  
\usepackage{color}
\usepackage{amsmath}
\usepackage{amssymb}
\usepackage{fancybox}

\begin{document}

\title{A Quantum Damper}
\author{Fumihiro Matsui} 
\affiliation{Department of Physics, College of Science and Engineering, Ritsumeikan University
Noji-higashi 1-1-1, Kusatsu 525-8577, Japan}
\author{Hiroaki S. Yamada}
\affiliation{Yamada Physics Research Laboratory,
Aoyama 5-7-14-205, Niigata 950-2002, Japan}
\author{Kensuke S. Ikeda}
\affiliation{College of Science and Engineering, Ritsumeikan University
Noji-higashi 1-1-1, Kusatsu 525-8577, Japan}

\date{\today}
\begin{abstract}
As an application of the classically decay-able correlation in a quantum chaos system 
maintained over an extremely long time-scale
 (Matsui et al; Europhys.Lett. {\bf 113} (2016) 40008),  
we propose a minimal model of quantum damper composed of a quantum harmonic 
oscillator  (HO) weakly interacting with a bounded quantum chaos system. 
Although the whole system obeys unitary evolution dynamics of 
only three quantum degrees of freedom, the mechanical work  applied to the HO is 
stationary converted linearly into the “internal energy” in time,
 characterized by an effective temperature in an irreversible way,  
 if the components of the quantum chaos system are mutually entangled enough. 
A paradoxical dependence of the duration time of the stationary energy conversion 
on the driving strength is also discussed.
\end{abstract}



\maketitle


\newcommand{\fracd}[2]{\frac{\displaystyle #1}{\displaystyle #2}}

\newcommand{\secondrev}[1]{\textcolor{red}{#1}}

\newcommand{\red}[1]{\textcolor{red}{#1}}
\newcommand{\firstref}[1]{\textcolor{red}{#1}}
\newcommand{\secondref}[1]{\textcolor{blue}{#1}}
\newcommand{\referee}[1]{\textcolor{magenta}{#1}}
\newcommand{\blue}[1]{\textcolor{blue}{#1}}
\definecolor{mygreen}{rgb}{0.2,0.8,0.2}
\newcommand{\green}[1]{\textcolor{mygreen}{#1}}

\def\tr#1{\mathord{\mathopen{{\vphantom{#1}}^t}#1}} 

\def\ni{\noindent}
\def\nn{\nonumber}
\def\bH{\begin{Huge}}
\def\eH{\end{Huge}}
\def\bL{\begin{Large}}
\def\eL{\end{Large}}
\def\bl{\begin{large}}
\def\el{\end{large}}
\def\beq{\begin{eqnarray}}
\def\eeq{\end{eqnarray}}

\def\eps{\epsilon}
\def\th{\theta}
\def\del{\delta}
\def\omg{\omega}

\def\e{{\rm e}}
\def\exp{{\rm exp}}
\def\arg{{\rm arg}}
\def\Im{{\rm Im}}
\def\Re{{\rm Re}}

\def\sup{\supset}
\def\sub{\subset}
\def\a{\cap}
\def\u{\cup}
\def\bks{\backslash}

\def\ovl{\overline}
\def\unl{\underline}

\def\rar{\rightarrow}
\def\Rar{\Rightarrow}
\def\lar{\leftarrow}
\def\Lar{\Leftarrow}
\def\bar{\leftrightarrow}
\def\Bar{\Leftrightarrow}

\def\pr{\partial}

\def\Bstar{\bL $\star$ \eL}

\def\>{\rangle}
\def\<{\langle}
\def\rr {\rangle\!\rangle} 
\def\ll {\langle\!\langle} 


\graphicspath{%
{./fig/}%
}

\def \eps { \varepsilon }
\def \H { \hat{H} }
\def \U { \hat{U} }
\def \Ukr { \hat{U}_{\rm KR} }
\def \p { \hat{p} }
\def \q { \hat{q} }
\def \I { \hat{I} }
\def \vIop{\hat{\bm {I}}}
\def \vI{\bm{I}} 
\def \th { \hat{\theta} }
\def \J { \hat{J} }
\def \ph { \hat{\phi} }
\def \aop { \hat{a} }
\def \bop { \hat{b} }
\def\Jop{\hat{J}}
\def \f { f }
\def \Hkqd { \H_{\rm{tot}} }
\def \Hho { \H_{\rm{HO}}(\q,\p) }
\def \Hkr { \H_{\rm{KR}}(\bm{\th},\bm{\I},t) }
\def \Vkr { {\cal V}(\bm{\th})}
\def \Hcat { \H_{\rm{cat}}(\th,\I,t) }
\def \Hacat { \H_{\rm{cat}}(\th_A,\I_A,t) }
\def \Hbcat { \H_{\rm{cat}}(\th_B,\I_B,t) }
\def \Hxcat { \H_{\rm{cat}}(\th_{\rm{dim}},\I_{\rm{dim}},t) }
\def \Hccat { \H_{\rm{cat2}}(\bm{\th},\bm{\I},t) }
\def \cat {ACM}
\def \stan {SM}
\def \Cr{Cr}
\def \rate {A}
\def \avrate {\overline{A}}
\def \Ndim {N_{\rm dim}}
\def\>{\rangle}
\def\<{\langle}
\def\stp{\tau}
\def\bbbZ{\Bbb Z}
\def\cT{\cal T}

\section{Introduction}  
How irreversible dynamics is self-organized inside of a small closed quantum system? 
What does the term ``thermalize'' mean in small closed quantum system? 
How small the closed thermalizable system can be made? 
Classical chaos can be the 
minimal origin of irreversibility \cite{prigogine}, 
because of the presence of mixing, which mean the decay of correlation, in a 
fully chaotic state. However, its quantum counterpart does not always play
the same role.

\begin{figure}[H] 
\begin{center}
 \centering
 \includegraphics[width=8cm]{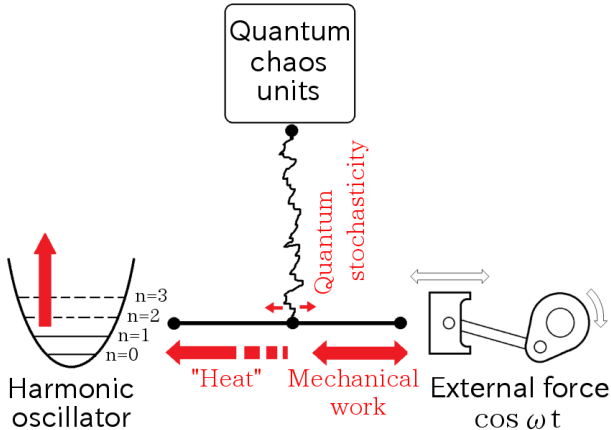}
 \caption{
A microscopic quantum damper constructed by combining 
quantum harmonic oscillator driven by a mechanical source
which is stochastically modulated by
a quantum chaos system.
}
 \label{Fig1}
\end{center}
\end{figure}

Consider an assembly of identical quantum chaos systems, which is interacting 
weakly with each other. 
Even if the system is unbounded and the 
associated Hilbert-space dimension $\Ndim$ is infinite, persistent quantum 
interference prevents time correlation from decaying completely, which means
microcanonical wandering over the phase space is suppressed, if the number of units (or
the dimension of the system) is less than a critical number 
 \cite{casati,fishman,casatianderson,delande,nandkishore}. 
On the other hand, in the case that $\Ndim$ is small, then even the number 
of units is large enough, the thermalization to equilibrium is often interrupted 
even though the system is non-integrable, as has been reported for quantum
many-body systems on the optical lattices \cite{kinoshita,mazets}.
The decay of correlation is quite delicate problem in the case of quantum system.

Many papers have been published for thermalization of isolated quantum
systems composed of many small quantum units (qubits) with a few quantum 
levels\cite{nandkishore,isolatedmanybodyquantum}. 
We, however, would like to explore in the opposite direction, namely, the irreversible 
relaxation in isolated quantum systems composed of a few large quantum units each of 
which has many quantum levels and further exhibits chaos 
in the classical limit \cite{ikedadissip,cohen}.
Irreversibility in small classically chaotic quantum system has been examined directly by 
numerical time-reversal experiments \cite{timerevexp,timerevexp-ballentine,benenti}. 
In particular, it has been extensively investigated by many investigators in the context of 
fidelity \cite{fidelity1,fidelity2}. 


In unbounded quantum chaos systems typically exemplified by
the standard map, the absence and presence of diffusion has been
of main interest in connection with the Anderson transition problem
\cite{fishman,delande}. 
The characteristics such as features of 
eigenspectrum \cite{wang09}, fidelity \cite{fidelity1,fidelity2} and so on has 
been investigated in the context of the Anderson transition.
On the other hand, we have been interested in the onset of time-irreversible 
behavior in finitely bounded chaotic quantum systems, which
is not realized as diffusion phenomena and is maintained only a 
limited time scale because of 
the finite Hilbert-dimension \cite{ikedadissip}.

In the previous paper we proposed a simple method which enables to investigate
the time-irreversible characteristics of the finitely bounded systems. 
With the method we can map the Fourier spectral features of correlation 
function to the diffusion characteristics of a fictious measurement system.
Applying the method to coupled quantum chaotic rotors,
we found that the time scale on which the decay-able quantum correlation is 
maintained is proportional to $\Ndim^2$,  not $\Ndim$ 
 if a full entanglement is achieved among the quantum chaos units \cite{matsui1,matsui2}. 
We called this ``lifetime'' of the decay-able correlation. 
The aim of the present paper is to demonstrate that 
by applying this extremely long lifetime of decay-able correlation in the quantum kicked
rotors we can design a most simple class of quantum damper 
(QD) which stationary converts the mechanical 
work into internal energy of the ``reservoir" over a sufficiently long time scale.

As will be introduced in Sec.\ref{sec:model}, 
our system is a quantum harmonic oscillator (HO) driven by periodic
force. It is coupled with coupled kicked rotors (KR) with bounded phase space, 
which exhibit ideally chaotic motion in classical limit. The chaotic KR perturbs
the coherent driving force stochastically which enables stationary
conversion of mechanical work into internal energy stored in the
HO in the classical limit. 
In Sects. \ref{sec:damper} and \ref{sec:damper-2}, 
we show that the development of entanglement in the coupled KR
makes the time scale on which the stationary irreversible energy 
transport is sustained very long time. However, the
actually observed time scale is much shorter than the theoretically
predicted one. The origin of such an unexpected result is 
considered in Sect.\ref{sec:coupling}.
We give the conclusion in Sect. \ref{sub:conclusion}.
Appendix \ref{app:derivation}
is devoted to exhibiting basic manipulations for exploring
energy transfer process in our system. 

\section{The Model}
\label{sec:model}
The main part of our system playing the role of the heat
reservoir is a quantum HO driven by
externally applied classical periodic force. The external
force excites the HO in a usual way, but the coupling of
the external force with the HO is parametrically modulated 
by a chaotic system, which makes the motion of HO
completely diffusive, thereby introducing an irreversible
nature. Figure 1 illustrates the proposed system 
in comparison with the classical damper. 
The total Hamiltonian reads as
\beq
  \Hkqd = \Hho+ \Hkr  + \eta \q f(\bm{\hat{I}}) \cos \omega t, 
\label{eq:totalhamiltonian}
\eeq
where $\Hho = \frac{\p^2}{2} + \Omega^2 \frac{\q^2}{2}$ is the Hamiltonian 
of HO with the frequency $\Omega$. The HO plays the role of the energy 
reservoir, which stores the energy transferred from the mechanical work done 
by the periodic force $\eta \cos \omega t$ in an ``irreversible'' manner.
$\Hkr$ indicate the Hamiltonian of classically chaotic quantum KR 
composed of two identical units interacting with weak coupling strength $\eps$ 
as
\begin{eqnarray} \label{eq:H_KR0}
&& \nn  \Hkr=\frac{\bm{\I}^2}{2T}+\Vkr\sum_{\ell=-\infty}^{\infty}\delta(t-\ell T)\\
&& {\rm with}~~~~~\Vkr =  \sum_{i=1}^2 K V(\th_i)+\eps V_{int}(\th_1,\th_2).
\end{eqnarray}
The angle operators $\th_i$ act in the bounded space $[0,2\pi]$, but 
the action operators $\I_i$'s space may be unbounded as is typically 
exemplified by the standard map. 
However, the chaotic region of normal dynamical systems is bounded 
and we are particularly interested in the effect of {\it boundedness}
of the phase space on the irreversible nature, and
we make the system bounded by imposing the periodic boundary conditions 
with the period $2\pi$
both on the action and the angle representation of wavefunction, respectively.
Hence the number of the quantum states $N$ 
of the unit system and Planck constant $\hbar$ are related by 
$2\pi \hbar N=2\pi\times 2\pi$, namely $\hbar=2\pi/N$.
$\eta$ is the strength of the external driving force of period $\omega$ 
working on the HO, and $f(\bm{\I})$ represents the interaction with the 
KRs which parametrically modulate the driving strength. 
It does not contains variables relevant for the kick operation.
We prepare the KR in  the classically fully chaotic state.

In executing the wavepacket propagation of the QD it is convenient to 
divide the time into the intervals of the kick period $T$ and to
construct the time-evolution unitary operator $\hat{U}_{\stp}$ from just after the 
$\stp$-th kick ($\stp\in\bbbZ$) to the $(\stp+1)$-th kick
by the product of two unitary evolution operators:
%
\begin{eqnarray}
\label{eq:quantummap}
&& \nn  \hat{U}_{\stp}=\hat{U}_{KR}\hat{U}_{{\rm HO},\tau}~~~~~{\rm with} \\
&& \nn  \hat{U}_{{\rm HO},\tau}=\\
&& \nn {\cal T} \exp\{-\frac{i}{\hbar} \int^{(\stp+1)T}_{\stp T} dt'\Hho + \eta\hat{q}f(\hat{\bm{I}})\cos\omega t'\}, \\
&& \hat{U}_{KR}=\exp\{-\frac{i}{\hbar}\Vkr\}\exp\{-\frac{i}{\hbar}\frac{\hat{{\bm I}}^2}{2}\}.
\end{eqnarray}
The time evolution starts just after the kick $t=\stp T+0$:
first the evolution by the driven HO and KR occurs for the period $T$, i.e.,applying 
$\e^{-\frac{i}{\hbar}\frac{{\bm \I}^2}{2}}{U}_{{\rm HO},\stp}={U}_{{\rm HO},\stp}\e^{-\frac{i}{\hbar}\frac{{\bm \I}^2}{2}}$, 
and next the KRs are kicked at $t=(\stp+1)T$, i.e., applying $\e^{-\frac{i}{\hbar}\Vkr}$. 
Technically a big merit to use HO is that the evolution by $\hat{U}_{{\rm HO},\stp}$ 
can be transformed into a product of three unitary operators each of which contains 
either $\p$ or $\q$ by the Baker-Campbell-Hausdorff expansion, which drastically
reduces the computation time.

In the following Sects. \ref{sec:damper} and \ref{sec:damper-2},
The coupling strength $\eta$ is taken such a small value that in the classical limit
the back-action of the HO to the KRs is negligibly small ($\eta \simeq O(10^{-4})$)
and the chaotic dynamics 
of the isolated KR is not disturbed by the coupling.
In Sect.\ref{sec:coupling}, the effect of  the coupling strength 
on the quantum dynamics is discussed.
%

\begin{figure*}[htbp]
\begin{center}
 \centering
 \includegraphics[height=5.0cm]{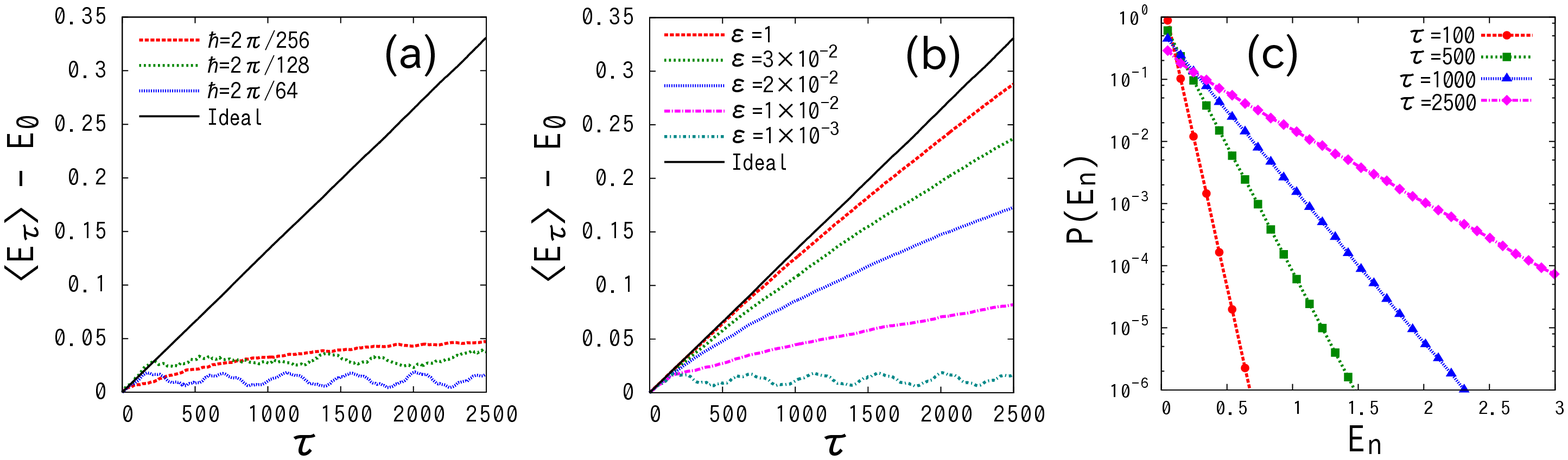} 
 \caption{
 (a)Absorbed energy $\<E_\tau\>-E_0$ by HO as a function of time
for $\hbar=2\pi/256,2\pi/128,2\pi/64$ in the single cat case ($\eps=0$).
(b)Absorbed energy $\<E_\tau\>-E_0$ by HO as a function of time
for various $\eps$'s with a relatively large $\hbar=2\pi/64$ 
in the twin cats case.
(c)Some snapshots of the energy distribution $P(E_n)$ 
of HO in the case $\eps=1$ and  $\hbar=2\pi/64$ in the panel (c).
All cases are $\eta=5 \times 10^{-4}$.
}
 \label{Fig2}
\end{center}
\end{figure*}

To be concrete, in what follows we take the coupled Arnold cat 
maps (ACM) as the KRs, 
namely, we choose $V(\th)=-\th^2/2$ and 
$V_{int}(\th_1,\th_2)=\cos(\th_1-\th_2)$ 
in Eq.(\ref{eq:H_KR0}). Further we couple the HO with only one of the two cats, i.e. 
$f(\hat{\bm{\I}})=\cos \I_1$, since we would like to restrict the path of influence by chaos.
We remark that the results shown in what follows are essentially the same 
if we choose another type of KR such as the  (unbounded) 
standard map $V(\th)=\cos\th$ when its classical version 
exhibits strongly chaotic motion.

\section{Operation of the damper}
\label{sec:damper}

\subsection{Energy transfer process}
We consider the time evolution of the absorbed energy by the HO.
With using the Heisenberg picture of the dynamics for annihilation and creation 
operators $\aop \equiv \sqrt{\Omega/2\hbar}(\q + i\p/\Omega)$ and $\aop^\dag$, 
or their envelope operators $\bop \equiv \aop \e^{i\Omega t}$ and $\bop^\dag$, 
one can obtain the following relation as shown in Appendix \ref{app:derivation}:
\begin{equation} \label{eq:bop}
 \bop_{\stp} - \bop \simeq -\frac{\eta (e^{i \nu T} - 1)}{2 \nu \sqrt{2 \hbar \Omega}} 
\sum^{\stp-1}_{k=0} e^{i k \nu T} \f_k,   
\end{equation}
where $\stp \in {\bf Z}$ and  
$\hat{X}_{s} \equiv \U_{1}^\dag\U_{2}^\dag..\U_s^\dag \hat{X}\U_{s}..\U_2\U_1$ 
indicate the Heisenberg operator just 
after the $s$-th kick.
$\nu=\Omega-\omega$ is the difference frequency, and we suppose 
the nearly resonant condition, namely, 
\beq
|\Omega-\omega| \ll 1 ~~{\rm and}~~~|\Omega-\omega|T\sim O(1).
\label{eq:resonant}
\eeq
 In our model there are two paths by which energy is exchanged
with external world. One is the periodic driving force and another is the kick force 
applied to KR. (Note that the energy of KR is finitely bounded.)
The nearly resonant condition mentioned above is very important to make 
the energy supplied to the HO come from the periodic driving force, 
and the exchange of energy between KR and HO is negligible
compared with the former as is shown in Appendix \ref{app:derivation}.

With this approximation the sum frequency term containing the frequency $\Omega+\omega$
 is ignored, and the energy stored in the HO is expressed by using the autocorrelation function
of the force due to the KR as follows:
\begin{equation} 
\label{eq:ene}
 E_\stp-E_0 = \sum_{s\leq \stp-1} \rate_s~~{\rm with}~~~
\rate_s=\mu\sum_{k = -s}^{s}\Cr_s(k)\e^{-i\nu k T},
\end{equation}
where $\mu\equiv\eta^2\frac{|\e^{i\nu T}-1|^2}{4\nu^2}$
and $\Cr_s(k)$ is the autocorrelation 
function of $\hat{f}_k\equiv f(\bm{\I}_k)$ defined by $\Cr_s(k)=\<\hat{f}_s\hat{f}_{s-k}\>$ 
for $k \geq 0$, and $\Cr_{s}(k)=\Cr_s(-|k|)^*$ for $k<0$.  (see Appendix \ref{app:derivation}.)
Its statistical property plays the crucial role in the energy transfer process.
In the classical limit the quantum average $\<..\>$ over 
the initial wavepacket is replaced by the average over
the ensemble of initial conditions having the same statistical weight in the 
phase space as the corresponding initial quantum wavepacket.

\subsection{Case of the single kicked rotor ($\eta \neq 0$, $\eps=0$)}
First, we consider the classical limit in the case of null coupling strength $\eps=0$, i.e., 
the HO is coupled with only one cat. 
The motion of the cat 
for a single step of the period $T$
is represented by the classical map 
$\theta_{\tau}=\theta_{\tau-1} + I_{\tau-1},~~I_{\tau}=K \theta_{\tau-1} + (K+1)I_{\tau-1}$, 
where $(\theta, I)$ are the classical canonical variables.
We are concerned with the hyperbolic case $K<-4,~0<K$~~($K \in \bbbZ)$ in which
the absolute value of Lyapunov exponent is larger than 1,  and the Lyapunov exponent
is the same everywhere in the bounded phase space $[0,2\pi]\times[0,2\pi]$,
where the periodic boundary condition is imposed as mentioned before. 
The cat map
exhibits mixing and so ergodic, and has a uniform invariant measure over the phase space for 
sufficiently large $K$. Further there remains no correlation between different steps
for almost all the initial distributions, 
\begin{equation} \label{eq:corr}
   \<f(\vI_{\tau})\>=0,~~~~~~~\<f(\vI_k)f(\vI_j)\>=\delta_{kj}\<f(\vI)^2\>, 
\end{equation}
if we take $f(\vIop_{\tau})=\cos I_1$.
All the above features provides a typical situation leading to a stationary energy transfer, 
and energy increases at the constant rate $A=\mu$ as,
\begin{eqnarray}
\label{eq:clsabsorption}
\<E_\stp\> - E_0 \simeq A\stp~~,~~~A=\mu\<f(\vIop)^2\>, 
\end{eqnarray}
because $\Cr_s(k)=\<f(\vI)^2\>\delta_{k,0}$.
We stress that, as is shown in Appendix, the absorbed energy is supplied by the 
periodic driving force, and not by the stochastic agitation by the KRs, 
if the nearly resonant condition is satisfied. 

Moreover,  Eq.(\ref{eq:bop}) is the summation of the complex random variable 
and so the distribution function 
for $\hat{b}_\tau$ obeys the complexified Gaussian process, 
which means the complex variables $(a,a^*)$ obey the Gaussian
distribution as
\beq
P(a^*,a) \propto \e^{-|a|^2/\<|a(t)|^2\>}.
\eeq
Accordingly, the distribution function of energy 
$E=\hbar \omega |a|^2$ becomes the exponential type 
$P(E) \propto e^{- E/\<E\>_\stp}$.

Next we consider its quantum version. 
In the quantum mechanical case we add
the cats a perturbation potential $\xi \cos (\th_{1,2} - \theta^{'}_{1,2})$ 
with classically negligible strength $\xi$ ($\sim O(\hbar^2)$), 
which suffices to break degeneracies 
caused by the underlying symmetries peculiar to the quantum 
ACM \cite{keating}.
The typical value of parameters we took are given in
the table \ref{parameters}.
The initial state is set to the direct product state $|\psi_0 \>=|0\> \otimes |I= \hbar N/2\>$, 
where $|0\>$ is the ground state of the HO,  and  $|I=\hbar N/2\>$ is the eigenstate of the
action operator $\I_i$ of the ACM.
 \begin{table}[H]
\begin{center}
 \caption{
\label{parameters}
The typical value of parameters we used in numerical calculation.
}
 \begin{tabular}{cc}
 \hline
{\rm parameters } & {\rm value} \\  \hline
$K $ & $10$ \\
$T $ & $10^2$ \\
$\Omega $ & $1$ \\
$\omega $ & $1+\sqrt{2}/T$ \\
$\xi $ & $4.25 \hbar^2$ \\
$\theta'_1 $ & $\sqrt{2}$ \\
$\theta'_2 $ & $-\sqrt{5}$ \\
\hline
$\eta $ & $5 \times 10^{-4}$ \\
$\hbar $ & $2\pi/N$ \\
$N $ & $2^{4} \sim 2^{8}$ \\
$\mu $ & $\sim 2.64 \times 10^{-4}$ \\
\hline
 \end{tabular}
\end{center}
 \end{table}

In the case of $\eps=0$, 
the time evolutions of absorbed energies by the HO,  
$\<E\>_{\tau}-E_0$, calculated numerically under the 
above conditions are plotted in Fig.\ref{Fig2}(a). The solid line in the figure represents 
theoretical values in the ideal classical limit. Dashed lines represent actual behavior for
$\hbar=2\pi/256$, $\hbar=2\pi/128$, and $\hbar=2\pi/64$ from the top.
As shown in Fig.\ref{Fig2}(a), energy absorbed by HO increases monotonously 
in the very initial stage but the increase is suppressed on a short time $\stp < O(10^2)$ 
and  turns into fluctuating behaviors around certain saturation levels in all cases.
The order of saturation levels in Fig.\ref{Fig2}(a) is about $O(10^{-2})$, 
which is the same order as $E_0$, and so we conclude that effective energy 
transfer were hardly realized in all cases.

As for the distribution function, it takes the Boltzmann type distribution only 
in the very early stage, but it becomes localized around the ground state 
whose shape is quite different from the exponential distribution. 
This is the reflection that the temporal energy absorption rate $\rate_\tau$ 
vanishes very promptly for $\eps=0$.

As has been discussed in detail in the previous paper \cite{matsui2}, 
the short saturation time means that in a single KR the number of 
eigenstates which are significantly connected with an eigenstate by 
the interaction potential $\eta q f(\hat{\bm{I}}) \cos\omega t$ 
in Eq.(\ref{eq:totalhamiltonian}) is much less than $N$. 
Such is very similar to the localization effect of eigenfunctions 
in unbounded isolated kicked rotor. Even though $N$ is increased more and 
the saturation time surely gets longer, but a recovery of stationary absorption
 can not be achieved. 
This is due to the strong quantum interference effect peculiar to 
low-dimensional quantum systems which has been observed as the ``scar''.\cite{scar}  
It corresponds to the Anderson localization
phenomenon \cite{casati, fishman} that has been observed widely 
in unbounded single KRs.

\subsection{Case of the coupled kicked rotors ($\eta \neq 0$, $\eps \neq 0$)}
Thus a single cat can hardly realize a QD. 
However, if the two cats are coupled, a remarkable entanglement transition 
occurs although the increment of the coupling strength $\eps$ is 
classically negligible and so very small \cite{matsui2,quantumclassical}
, which is a well-known phenomenon in coupled quantum 
chaos systems \cite{couplerotorentangle}.      

Then the effective dimension $\Ndim$ of the Hilbert space
of the cats increases from $N$ to $N^2$.
Let us define the time-scale $\tau_{QD}$ of saturation of 
the stationary energy absorption caused by the quantum suppression of 
the classical decay of autocorrelation.
It is expected to increase markedly by the coupling between the cats.
Indeed, in our previous paper we showed that the finite time 
Fourier transform of  the quantum autocorrelation function
\beq
\label{eq:autcorr}
F_\stp(z) \equiv \sum_{k=-\stp}^\stp\Cr_\stp(k)\e^{-izTk}, 
\eeq
which is nothing more than the temporal absorption rate $\rate_\stp$ for 
$z=\nu =(\Omega-\omega)$
in Eq.(\ref{eq:ene}), 
takes the constant classical value due to the classical chaotic decay of 
autocorrelation function if $\stp$ is less than the ``lifetime'' $\tau_L$ \cite{matsui1}:
\beq
\label{eq:lifetime}
 \tau_L \sim 
\begin{cases}
  C\Ndim^2=CN^4 &  \text{for~~} z\neq 0\\
  C\Ndim=CN^2 &  \text{for~~} z = 0, 
\end{cases}
\eeq
where the coefficient $C(z)$ is given as
\beq
C(z) \sim \Cr_{cl}(0)/\sum_{s=-\infty}^\infty\Cr_{cl}(s)\e^{-izTs}
\eeq
with the classical correlation function $\Cr_{cl}(s)$.
For $\tau>\tau_L$, 
$F_\stp(z)$ decays to zero due to the multiple-periodic nature of 
the quantum correlation of the bounded system.
%
The Eq.(\ref{eq:lifetime}) becomes crucial again in Sect.\ref{sec:coupling}, 
and we will explain the physical origins there.

 If we can identify $\tau_{QD}$ with $\tau_L$, the absorption rate $\rate_\stp$
 keeps the classical rate over an extremely long time-scale of $\tau_L$, 
and the stationary and one way conversion of the 
mechanical energy to the ``thermal'' energy is almost accomplished.

The fully quantum result of the time-dependent energy absorption by the HO are 
depicted in Fig.\ref{Fig2}(b).  Results for different values of $\eps$ are plotted,  
where $\hbar=2\pi/64$ and $E_0 = \hbar\Omega/2 \simeq 6.9 \times 10^{-2}$ 
for all results.
The line running close to the bottom is the result of $\eps=0$, while the lines 
running in the upper sides are those for significantly larger values of $\eps$.
The solid line represents the ideal classical behavior expected 
by Eq.(\ref{eq:ene}).

It is evident that the quantum behavior approach to the ideal classical one as 
the value of $\eps$ increases. The absorbed energy at $\tau=2500$ for a sufficiently 
large coupling strength $\eps=1$ is about $3.4 \times 10^{-1}$, which is 5 times larger 
than $E_0$. This implies that a one-way transfer of the energy from the driving source
to the HO occurs if $\eps$ is taken as a sufficiently large value.

Figure \ref{Fig2}(c) depicts the time evolution of energy distributions 
for a relatively large value of $\eps$ in the examples in Fig.\ref{Fig2}(b),
i.e.  $\eps=1, \hbar=2\pi/64$.
Exponential distributions, as expected in the ideal case,  
are certainly realized and are sustained for a long time scale.
Namely, it follows very well that the distribution becomes 
the Maxwell-Boltzmann type one at least up to the attainable time by 
the numerical simulation. 
The effective temperature is represented by $E_\stp$, which 
increases almost linearly in time as is shown in Fig.\ref{Fig2}(b).

\section{$\hbar-$dependence of the quantum damper}
\label{sec:damper-2}
\begin{figure*}[htbp]
\begin{center}
 \includegraphics[height=5.0cm]{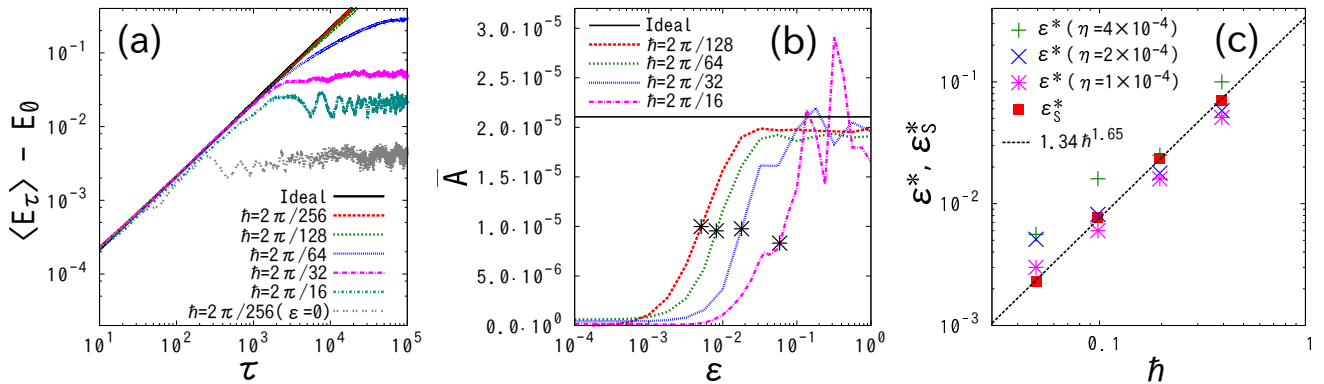}
 \caption{
Energy absorption in the case of coupled twin cats.
(a)Long-time behavior of the absorbed energy $\<E_\tau\>-E_0$ for 
decreasing $\hbar$s, where $\hbar=2\pi/N$ with $N=2^{4},2^{5},2^{6},2^{7}$. 
Here, $\eps=1$ and  $\eta=2 \times 10^{-4}$. 
(b)Average absorption rate $\overline{A}$ as a function of $\eps$ 
for various $\hbar=2\pi/N$s. $\eta=2 \times 10^{-4}$. 
The horizontal line shows absorption rate in classical theory.
(c)The threshold values $\eps^{*}$ at some values of $\eta$ and $\eps^{*}_S$ 
as a function of $\hbar$.
The reference line denotes $\eps^{*}=1.3 \hbar^{1.65}$,
which is fitted in the range $\tau \in \left[0,\tau_{av}\right]$. 
}
\label{Fig3}
\end{center}
\end{figure*}

As is expected by Eq.(\ref{eq:lifetime}), we demonstrate in Fig.\ref{Fig3}(a) that the 
time of the saturation $\tau_{QD}$ until which the stationary absorption mimics the 
ideal classical absorption process gets markedly longer as the Planck constant 
$\hbar=2\pi/N$ is reduced by increasing the dimension $N$. 
On the other hand, for fixed $\hbar$, the long-lived stationary 
absorption process is realized beyond a threshold of the coupling strength $\eps$.
We investigate here the $\hbar-$dependence of 
such a characteristic transition phenomenon.

The change of the absorption process with the increase in $\eps$
is qualitatively captured by the variation of average absorption rate  $\avrate$.
We define the transient
stationary energy absorption rate $\avrate$ 
by the least square fit of the data to  $\< E_\stp \>-E_0=\avrate\stp$
for a fixed  interval $t \in \left[0,\tau_{av} \right]$, 
where $\tau_{av}=0.5\tau_{QD}$ \cite{tav}.

We show in Fig.\ref{Fig3} (b) the average energy absorption rate $\avrate$ as a 
function of $\eps$ for three values of $\hbar$ as $\hbar=2\pi/2^4, 2\pi/2^5, 2\pi/2^6$.
We set $\eta=2.0 \times 10^{-4}$  for all cases.
As shown in the figure it seems that the rate suddenly increases as $\eps$ exceeds a certain 
threshold value $\eps=\eps^*$ and beyond it the $\avrate$ reaches to a plateau along which
the rate does not change significantly. The plateau height is close to the classical absorption 
rate for all $\hbar$ although some significant ``quantum fluctuation'' is overlapped for 
relatively large $\hbar=2\pi/16$. We may, therefore, call the plateau as the ``classical plateau''. 
Moreover, it is evident that the plateau and the threshold $\eps^*$ shifts 
to the smaller side of $\eps$ as $\hbar$ decreases, which becomes 
more evident by the plots $(\hbar,\eps^*)$ depicted in Fig.\ref{Fig3}(c).

The approach of the average absorption rate to the classical value means
the remarkable increase of $\tau_{QD}$, which should reflect the rapid
growth of entanglement between the two cats, as was suggested in the previous
section. It means the number of eigenstates connected strongly by the interaction
potential $\eta f(\hat{\bm{I}})\cos\omega t$ increases with increase in the coupling strength 
$\eps$ \cite{matsui1,matsui2}.
 Indeed, the increase in the number of the connected 
eigenstates via the operator $f(\hat{{\bm I}})=\cos \hat{I}_1$ contained
in the interaction potential is possible only by 
the development of entanglement between the two cats,
which should be directly observed by the entanglement entropy (EE).

The approach of the average absorption rate to the classical value means
the remarkable increase of $\tau_{QD}$, which should reflect the rapid
growth of entanglement between the two cats, as was suggested in the previous
section. It means the number of eigenstates connected strongly by the interaction
potential $\eta f(\hat{\bm{I}})\cos\omega t$ increases with increase in the coupling strength 
$\eps$ \cite{matsui1,matsui2}.
 Indeed, the increase in the number of the connected 
eigenstates via the operator $f(\hat{{\bm I}})=\cos \hat{I}_1$ contained
in the interaction potential is possible only by 
the development of entanglement between the two cats,
which should be directly observed by the entanglement entropy (EE) 

Indeed, the increase in the number of the connected 
eigenstates via the operator $f(\hat{{\bm I}})=\cos \hat{I}_1$ contained
in the interaction potential is possible only by 
the development of entanglement between the two cats,
which should be directly observed by the entanglement entropy (EE).

We follow Ref.\cite{matsui2}, and introduce the EE
defined for every eigenstate $|n\>$ of the evolution 
operator of the coupled cat $\U_{\rm KR}$, i.e.,
\beq
S_n \equiv -{\rm Tr} |n><n| \log |n\>\<n|
\eeq
where trace is taken for either of the twin cats.
$S_n$ varies from $0$ to $\log N$, and $S_n=0$ indicates no entanglement happens. Next
the mean EE averaged over all the eigenstates, i.e., $S \equiv \sum_n S_n / N^2$ is introduced,
and the $\eps$-dependence of $S$ is explored, which draw characteristic curves similar to 
Fig.\ref{Fig3}(b) starting with $S=0$, and steeply increases in the regime close 
to $\eps \sim \eps^*$. \cite{matsui2}

To analyse $\hbar$-dependence of the threshold of the mean EE, 
we define the threshold value $\eps^*_S$ satisfying the relationship 
$S(\eps_S^*) = S(\eps=1)/2$, and
plotted $\eps_S^*$ versus $\hbar$ in Fig.\ref{Fig3}(c) by rectangle marks.
As shown in the figure, the points of $(\hbar,\eps^*_S)$ agree very well with the threshold of
absorption $(\hbar,\eps^*)$ and they are aligned very well along the dashed line 
which indicates 
 $\eps^{*} \propto \hbar^{\alpha}$ with 
$\alpha = 1.65$. This fact is a direct evidence that the enhancement
of entanglement happening in the isolated coupled KR is the origin of the increase in the 
stationary absorption rate of the QD.

\section{Duration time of the stationary conversion}
\label{sec:coupling}

In this section,  we investigate the $\eta-$dependence of the 
duration time $\tau_{QD}$ and give the interpretation.

\subsection{$\eta-$dependence }
\begin{figure}[H] 
 \centering
 \includegraphics[height=13.0cm]{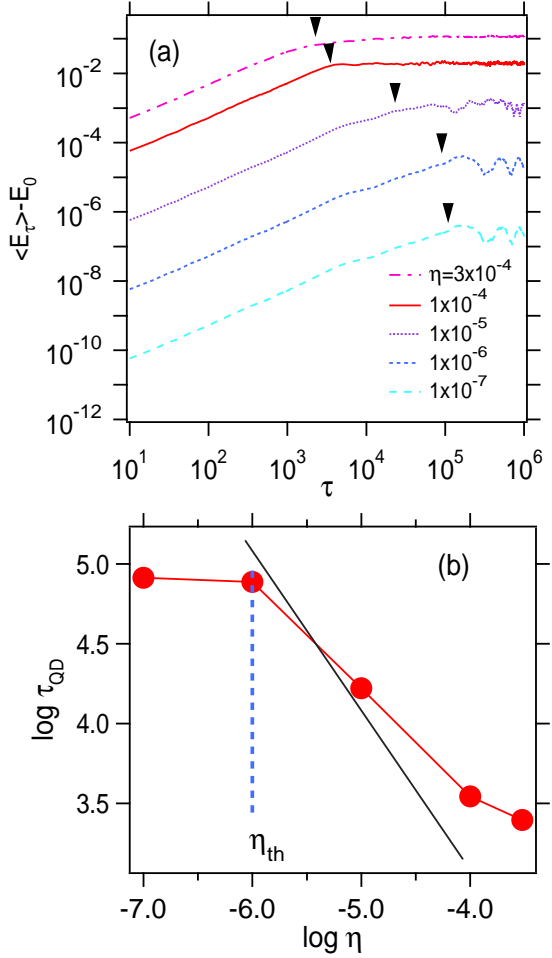}
 \caption{
(a)Long-time behavior of the absorbed energy $\<E_\tau\>-E_0$ of HO
at various values of $\eta$ for the twin cats ($\eps=1$), where
$\hbar=2\pi/32$. The arrow indicates $\tau_{QD}$. 
(b)The $\eta$ dependence of $\tau_{QD}$, which is compared with
$\tau_{QD} \propto \eta^{-1}$ denoted by dashed line. 
The location of $\eta_{th}$ is also indicated. 
}
\label{Fig4}
\end{figure}

Finally, let us focus on the fully entangled regime $\eps\gg\eps^*$
of the two cats and demonstrate how the stationary flow of
the energy absorption process by the QD is strengthen
with the increase in the Hilbert space dimension $\Ndim =
N^2$ of the coupled cats. As shown in Fig.3(a), it is evident
that with increase in $\Ndim$ , the time of saturation of the
energy absorption markedly increases. i.e., $\tau_{QD}\sim 10^3$ for
$\Ndim = 16^2$ , $\tau_{QD}\sim 5\times10^3$ for $\Ndim = 32^2$, and
$\tau_{QD}\sim 2\times10^4$ for $\Ndim = 64^2$.
For $\Ndim > 128^2$ the absorption curve closely follows 
the ideal curve up to the time numerical simulation is 
attainable.
However the time scale of saturation seems to be much
shorter than the prediction of $\tau_L\sim N^4$ in Eq.(\ref{eq:lifetime}) for
$z = \nu \neq 0$. 
The Eq.(\ref{eq:lifetime}) holds under the condition that 
the coupled cats are completely isolated, i.e. $\eta \to 0$.
That is, the difference between the $\tau_L$ and $\tau_{QD}$ 
should be due to be the interaction of the coupled cats with HO and the
driving force.
The increase of $\eta$ changes the condition ($\eta \to 0$)
very significantly. Indeed, in Fig.\ref{Fig4} we show how the 
absorption curve changes as $\eta$ is varied, where $\Ndim=32^2$.  
It is evident that the time of saturation $\tau_{QD}$ is 
lengthen as $\eta$ is reduced, and there is a threshold $\eta_{th}$ of $\eta$ 
below which the absorption curves are only parallel downward shifts
and thus $\tau_{QD}$ takes the maximum constant value $\tau_{QD} \sim 10^5$. 
which should $\tau_L$ in Eq.(\ref{eq:lifetime}).

\subsection{Isolated KR again}
It seems quite strange that the coupling with the HO together 
with the driving source, which should result in an effective 
increase of associated Hilbert dimension, reduces $\tau_{QD}$.
This very paradoxical phenomenon can be qualitatively
explained following the arguments of deriving the expression of 
$\tau_L$ in Eq.(\ref{eq:lifetime}).
(For the more detailed arguments see Eqs.(10) and (11) 
of the Ref.\cite{matsui1}.)
We give a rough sketch as follows. Let $\gamma_j$
be the eigenangle of the unitary evolution operator $\U_{KR}$
of the coupled cats. Since the time evolution of the pair
correlation function is described by the paired eigenangles,
say $\gamma_{j}$ and $\gamma_{j'}$ via the oscillating factor 
$e^{-i(\gamma_j-\gamma_{j'})\tau}$, where the summation is taken over
all $\Ndim$ eigenstates because the cats completely chaotic
and the perturbation connects all the states.
Thus the saturation of $F(z=0)$ occurs at the time-scale on which
the contribution from the term having the minimal 
difference of eigenangle $\delta={\rm Min}_{j,j'}|\delta_{j,j'}|$
,where $\delta_{j,j'}=\gamma_j-\gamma_{j'}$, becomes oscillatory,
namely $\tau_L\delta\sim O(2\pi)$. Since the strong level repulsion of
entangled chaotic cats yields $\del\sim 1/\Ndim = 1/N^2$, we
conclude that $\tau_L\sim \Ndim = N^2$, which is $z = 0$ case of 
Eq.(\ref{eq:lifetime}). This time-scale coincides with the
so-called Heisenberg time.

On the other hand, for $z=\nu \neq 0$ the additional frequency
changes the oscillatory factor as $e^{-i(\gamma_j\pm zT-\gamma_{j'})\tau}$ 
and the above rule is modified completely: the minimal 
difference　$\delta={\rm Min}_{j,j'}|\delta_{jj'}|$, where 
$\delta_{jj'}\equiv |\gamma_j\pm zT-\gamma_{j'}|$, 
decides $\tau_L$. The frequency difference $\delta_{jj'}$ can be interpreted
as the difference in the paired eigenangles of eigenstates in the extended space
formed by the direct product of the cats, the HO, and the driving source.
The emergence of $z(=\nu)$ in the frequency difference $\delta_{jj'}$ 
makes the choice of the minimal $|\delta_{jj'}|$ free from the presence of 
level repulsion among $\gamma_j$s, and so $\delta$ can be much smaller 
than the case of $z = 0$, and the minimal value reaches 
to $\delta \sim 1/\Ndim=N^{-4}$, which leads to the case of $z\neq0$ of 
Eq.(\ref{eq:lifetime}) \cite{matsui1}.

\subsection{Entanglement between HO and ACM}
However, the growth of the interaction among the coupled cats, 
external driving force and HO
make any closely approaching pair of quasi-energies, say, $\gamma_j\pm zT$ 
and $\gamma_{j'}$ to repulse
by introducing the coupling between the paired eigenstates
in the extended space mentioned above, and the minimal scale  
$\delta \sim 1/\Ndim=N^{-4}$ is no longer maintained.

With the above arguments, we can understand the behavior in 
Fig.\ref{Fig4} very roughly as follows: as the coupling strength 
$\eta$ become greater than the minimal spacing $\sim 1/N_{dim}^2$, then the 
quasi-degenerate pair of energy levels $\delta_{jj'}$ repulses with each other 
to separate in proportional to the perturbation strength $\eta$, which couples 
the KR with other degrees of freedom. That is, 
\beq
\label{etath}
  \eta_{th} \simeq \frac{1}{\Ndim^2}.
\eeq
In such a limit the saturation level of absorption, which is evaluated by summation over
the temporal absorption rate Eq.(\ref{eq:autcorr}), is dominated by the minimal
scale of the energy-level spacing  $\delta_{jj'} \sim \eta $ as
\beq
    \eta^2 \times \sum_{j,j'} (N\delta_{jj'})^{-2} \propto \eta^2 \times \eta/\eta^2 \sim \eta 
\eeq 
because the number of the eigenstates $j'$ less than $\delta_{jj'}<\eta$ 
is roughly given by $\eta/N^{-2}=\eta N^2$ for each $j$. 
On the other hand, the absorped energy classically increases as  
$\propto \eta^2 t$ according to Eq.(\ref{eq:clsabsorption}), and so $\tau_{QD}$ is
given by
\beq
\label{tauQD}
    \tau_{QD} \propto \eta^{-1}.
\eeq 
Figure \ref{Fig4} indicates that $\tau_{QD}$ starts to decrease from 
$\eta_{th}$ given by Eq(\ref{etath})($N=32$) for sufficiently small $\eta$, 
and further decreases very according to Eq.(\ref{tauQD}).
The decrease continues with $\eta$ up to $\tau_{QD} \sim \Ndim$, 
namely the case of $z = 0$ in Eq.(\ref{eq:lifetime}).

It should be noted that the $\Ndim$ is no longer $N^2$ but $N^2\times N_R$ , 
where $N_R$ being the mean number of the states of HO and the driving source 
which are entangled with the coupled cats
Detailed analyses for the above rather rough arguments will 
be presented for more simple standard model proposed in \cite{matsui1}.

As long as both HO and the driving source are outsider of the cats group, 
 an extremely long lifetime is promised. But if they are 
taken into the cats group 
as the internal degrees of freedom
they lose the extremely long lifetime in compensation 
for the large transfer rate.

\section{Conclusion}
\label{sub:conclusion}
Based upon the results of previous works, we proposed a very simple
fully quantum mechanical model which can transform the mechanical energy into
the internal energy in an irreversible way like a classical damper.
Our system is composed of only
three degrees of freedom: one is a quantum harmonic oscillator (HO)
playing the role of the energy storage and another two are quantum kicked rotors
(KR) which are classically chaotic. The KRs are defined in bounded phase space
and so their Hilbert spaces has finite dimensions. 
The HO is driven by a coherent periodic force and KRs 
work as a stochastic source disturbing the phase of the driving force,
and one can show an irreversible stationary energy transfer from the driving
source to HO is realized in the ideal classical limit. In quantum mechanics, however, 
the irreversible stationary energy transfer continues up to a finite time scale
denoted by $\tau_{QD}$.

The time scale $\tau_{QD}$ is expected to be sensitively dependent upon the coupling 
strength $\eps$ between the KRs, and it increases very rapidly
as $\eps$ exceeds a very weak threshold decided by the Planck constant.
The increment of $\tau_{QD}$ is closely correlated with
the development of entanglement between the two KRs with increase in $\eps$.
However, the maximal $\tau_{QD}$ realized after the entanglement was much shorter 
than the one predicted by the previous theory for the isolated chaotic KRs.

The reduction of $\tau_{QD}$ is caused by the entanglement of the KRs with other
systems i.e. HO and the driving source. The entanglement removes the quasi-degeneracy
of the eigenstates of the whole system, which was the origin of very large $\tau_{QD}$.  
The enhanced entanglement in the coupled KR enhances $\tau_{QD}$, whereas it makes the 
threshold $\eta_{th}$ of the coupling strength $\eta$ of KR with HO and driving source
extremely small. Development of entanglement in the whole system works inversely and 
reduces the time scale predicted in the ideal limit much shorter.
Thus the development of the quantum entanglement plays opposite roles in realizing 
the time-irreversible behavior. A complete understanding of the relation between 
the time scale of quantum irreversibility $\tau_{QD}$ and the development of
entanglement among the quantum elements is still an open problem,
and a further clarification in a more simple and typical situation is strongly desired. 

\appendix

\section{Analytical derivations by balance equation}
\label{app:derivation}
%
We summarize here some basic calculations necessary for several equations
used in the present paper. We also discuss here the balance between
the absorbed energy by the HO and the energy supplied by the periodic driving source.

With using the Heisenberg picture of the dynamics for annihilation and creation 
operators $\aop \equiv \sqrt{\Omega/2\hbar}(\q + i\p/\Omega)$ and $\aop^\dag$,
one can immediately obtain

%
\begin{equation} 
\label{eq:aopEM}
     \frac{d\aop}{dt} = -i\Omega \aop - i\eta f(\hat{\bm {I}})\cos(\omega t), 
\end{equation}
which is easily integrated by introducing the slowly varying envelope operator 
$\aop = \e^{-i\Omega t}\bop$ as, 
\begin{eqnarray} 
\label{eq:bop0}
\nn \Delta \bop(t)& :=& \bop(t)-\bop(T\tau)  \\
&=& -\frac{1}{2\sqrt{2\Omega\hbar}}f(\vIop_0)\times  \nonumber \\
\nn && \biggl[\frac{e^{i(\Omega-\omega)t}-e^{i(\Omega-\omega)T\tau}}{\Omega-\omega}
                                      +\frac{e^{i(\Omega+\omega)t}-e^{i(\Omega+\omega)T\tau}}{\Omega+\omega}\biggr] \\
   &\simeq &  -\frac{1}{2\sqrt{2\Omega\hbar}}f(\vIop_0)\frac{e^{i(\Omega-\omega)t}-e^{i(\Omega-\omega)T\tau}}{\Omega-\omega}
\end{eqnarray}
for $T\tau \leq t \leq T(\tau+1)$ under the nearly resonant limit such that 
$(\Omega-\omega)T=\nu T=C\sim O(1)$.
Using this relation iteratively from $t=0$ to $t=\stp T$, we obtain Eq.(\ref{eq:bop}). 
By using  Eq.(\ref{eq:bop}) and the definition
\begin{eqnarray} 
\label{eq:qbop}
    \q(t)=\sqrt{2\hbar}{\Omega}[\bop(t)e^{-i\Omega t}+h.c.], 
\end{eqnarray}
the energy stored by the HO at the step $\tau$ is given by
\begin{equation} 
\label{eq:aop}
\<\aop^\dag_{\stp}\aop_{\stp}\> = 
\mu \sum_{k=0}^{\tau-1}\sum_{j=0}^{\tau-1}\e^{i\nu(k-j)}\<f(\vIop_k)f(\vIop_j)\>, 
\end{equation}
which is rewritten as Eq.(\ref{eq:ene}).

Now we suppose that the KR follows chaotically ideal behavior realized in the classical limit, 
and consider the energy balance between the driving source and HO.
We suppose that the coupling strength $\eta$ among HO, the driving force and KRs is 
so weak that the correlation characteristics of KR represented by Eq.(\ref{eq:corr}) holds
in the classical limit.  Then the energy of HO increases linearly 
obeying Eq.(\ref{eq:clsabsorption}) at the classical absorption rate per step
\begin{equation} \label{eq:HOene}
          A = \eta^2\<f(\vI)^2\>\frac{1-\cos(\Omega-\omega)T}{4(\Omega-\omega)^2}.
\end{equation}
Next, we evaluate the energy transfer from the driving source during a single period 
$\tau T< t  \leq (\tau+1)T$.
To this purpose we extend our QD Hamiltonian so as to include the degree 
of freedom of driving source:
let $\omega \Jop$ be the energy of driving source, where $\Jop=-id/d\phi$ is the operator
of action variable representing the driving source and $\phi$ is the angle variable describing
the phase of the driving source. Then the interaction Hamiltonian 
$\eta f(\vIop)\cos(\omega t)q$ 
in Eq.(\ref{eq:totalhamiltonian}) should be replaced as  $\eta f(\vIop)\cos(\phi)q$, 
and the extended total Hamiltonian including the driving source is written as, 
\begin{equation} 
\label{eq:extendH}
  H_{ext} =  \Hho+ \Hkr  + \eta \q f(\bm{\hat{I}}) \cos \phi + \omega \Jop.
\end{equation}
%
The Heisenberg equations of motion for the variables of driving source
are the same as the canonical equation of motion
\beq
\label{eq:extendeq}
  \frac{d\phi}{dt} &=& \frac{i}{\hbar} [H_{ext},\phi ] = \omega,  \\    
  \frac{d\Jop}{dt} &=& \frac{i}{\hbar} [H_{ext},\Jop ]  = \eta \q f(\vIop)\sin\phi.
\eeq
By integtating Eq.(\ref{eq:extendeq}) from $(\tau-1)T$ to $\tau T$, 
the variation of the source energy $B=\omega(J(T(\tau+1))-J(\tau T))$ should be
\begin{equation} 
\label{eq:Jene0}
 B = \eta \int_{\tau T}^{(\tau+1)T} \<f(\vIop_\tau) \q(t)\>\sin\omega t dt.
\end{equation}

From Eqs.(\ref{eq:qbop}) and (\ref{eq:bop0}),  
$q(t)$ is the sum of the terms proportional to $\bop(\tau T)$ 
and to $\Delta \bop(t)$. 
The contribution from the former to the r.h.s. of Eq.(\ref{eq:Jene0}) vanishes
because it is the sum of the terms proportional 
to $f(\vIop_j)$ with $j<\tau$ from Eq.(\ref{eq:bop}). 
The contribution comes only from $\Delta \bop(t)$, 
and in the nearly resonant condition Eq.(\ref{eq:corr}), it leads to the result 
\begin{equation} 
   B = \frac{\omega}{\Omega}\eta^2\<f(I)^2\> 
\frac{\cos((\Omega-\omega)T)-1}{4(\Omega-\omega)^2}, 
\end{equation}
 which coincides with the absorption rate $A$ of Eq.(\ref{eq:HOene})
 in the nearly resonance condition.  



\section*{Acknowledgments}
This work is partly supported by Japanese people's tax
via JPSJ KAKENHI 15H03701, and the authors would
like to acknowledge them. They are also very grateful to
Kankikai, and Koike memorial house for use of the facilities during this study.




\begin{thebibliography}{00}
\bibitem{tav}
The classical absorption process is clearly recovered at least in the the time range 
$0 \leq \tau \ll \tau_{QD}(\eps)$ at the classical coupling strength $\eps\sim O(1)$. 
Then we take $\tau_{av} = 0.5 \times \tau_{QD}(\eps=1)$ as the common time scale on 
which the least square fit is done.
\end{thebibliography}
\end{document}